\documentclass[final,authoryear,1p]{elsarticle}
\pdfoutput=1
\usepackage{graphics}
\usepackage{epsfig}
\usepackage{bm}
\usepackage{amssymb,amsmath}
\usepackage{longtable}
\usepackage{subfig}
\usepackage{multicol}
\usepackage{supertabular}
\usepackage{bookmark}% http://ctan.org/pkg/bookmark
\usepackage{courier}
\usepackage{comment}
\usepackage{listings}
\usepackage{xcolor}
\usepackage{hyperref} %enabling hyper references
\usepackage{url} % reference by url
\hypersetup{colorlinks=true}
\begin{document}

\definecolor{mygreen}{rgb}{0,0.6,0}
\definecolor{mygray}{rgb}{0.5,0.5,0.5}
\definecolor{mymauve}{rgb}{0.58,0,0.82}
\lstset{
  backgroundcolor=\color{white},   % choose the background color; you must add \usepackage{color} or \usepackage{xcolor}
  basicstyle=\footnotesize\ttfamily,        % the size of the fonts that are used for the code
  breakatwhitespace=false,         % sets if automatic breaks should only happen at whitespace
  breaklines=true,                 % sets automatic line breaking
  captionpos=b,                    % sets the caption-position to bottom
  commentstyle=\color{mygreen},    % comment style
  deletekeywords={...},            % if you want to delete keywords from the given language
  escapeinside={\%*}{*)},          % if you want to add LaTeX within your code
  extendedchars=true,              % lets you use non-ASCII characters; for 8-bits encodings only, does not work with UTF-8
  frame=single,	                   % adds a frame around the code
  keepspaces=true,                 % keeps spaces in text, useful for keeping indentation of code (possibly needs columns=flexible)
  keywordstyle=\color{blue},       % keyword style
  language=C++,                    % the language of the code
  otherkeywords={*,...},           % if you want to add more keywords to the set
  numbers=left,                    % where to put the line-numbers; possible values are (none, left, right)
  numbersep=5pt,                   % how far the line-numbers are from the code
  numberstyle=\tiny\color{mygray}, % the style that is used for the line-numbers
  rulecolor=\color{black},         % if not set, the frame-color may be changed on line-breaks within not-black text (e.g. comments (green here))
  showspaces=false,                % show spaces everywhere adding particular underscores; it overrides 'showstringspaces'
  showstringspaces=false,          % underline spaces within strings only
  showtabs=false,                  % show tabs within strings adding particular underscores
  stepnumber=1,                    % the step between two line-numbers. If it's 1, each line will be numbered
  stringstyle=\color{mymauve},     % string literal style
  tabsize=2,	                   % sets default tabsize to 2 spaces
  title=\lstname                   % show the filename of files included with \lstinputlisting; also try caption instead of title
}

\begin{frontmatter}
\title{Implementation and validation of two-phase boiling flow models in OpenFOAM}
%\title{This is a specimen title\tnoteref{t1}}
%\tnotetext[t1]{This document is a collaborative effort.}

\author[NRT]{Kai Fu}
\ead{kaifu@kth.se}
\author[NRT]{Henryk Anglart\corref{cor1}}
\ead{henryk@kth.se}
\address[NRT]{Division of Nuclear Reactor Technology, Royal Institute of Technology, S-106 91 Stockholm, Sweden}

\let\thefootnote\relax\footnote{Abbreviations: DNB, departure from nucleate boiling.}
\cortext[cor1]{Corresponding author. Tel.: +46-8-5537-8887}

%Highlights:
%The interfacial area concentration transport model has been implemented in OpenFOAM.
%The Bartolomej and DEBORA experiments were selected as the validation case.
%The sensitivity of turbulent dispersion force coefficient was tested.
%The prediction of void fraction and the liquid temperature is quite satisfied.
%\input epsf

\begin{abstract}
Prediction of two-phase boiling flows using the computational fluid dynamics (CFD) approach is very challenging since several sub-models for interfacial mass, momentum and energy transfer in such flows are still not well established and require further development and validation. Once validating a particular model, it is important that all key parameter involved in the model are carefully verified. Such verification is typically performed by separate effect tests, where one parameter at a time is compared to a measured or otherwise known value. Needless to say that for complex models, which are typical for CFD applications to two-phase flow, the number of independent parameters that need to be verified can be quite high. This particular feature makes the validation process of complex CFD models in open source codes very attractive, since full access to the implementation details is possible.

This paper is concerned with implementation and validation of two-phase boiling bubbly flow models using the OpenFOAM, open source environment. The model employs the two-fluid formulation of the conservation equations with the Reynolds-averaged treatment of the turbulent terms. The model consists of six conservation equations for the liquid and the vapor phase, allowing for the thermodynamic non-equilibrium and compressibility of both phases. In addition, the model includes two transport equations for the turbulence kinetic energy and energy dissipation and one transport equation for the interfacial area concentration. New models for wall heat partitioning as well as for the phase change terms in nucleate boiling have been implemented. Sensitivity studies as well as validation of the model against measured data available in the open literature have been performed and it has been shown that a reasonable agreement between predictions and experiments has been achieved.
\end{abstract}


\end{frontmatter}

\section{Introduction}
%Subcooled flow boiling is a stage which is between single phase flow and saturation flow boiling. In subcooled boiling, regions of bulk boils and the fluid travels through some narrow channels. The formation of large bubbles could block the passage of the fluid and results in a departure from nucleate boiling (DNB) which means the bubbles always dominate over the solid surface and a dry patch appears. The dry patch on the surface could isolate the fluid from the surface and thus the heat transfer coefficient from surface to the bulk is greatly reduced. That is boiling crisis in subcooled boiling flows. In this subject a mechanistic model of critical heat flux under subcooled flow boiling conditions was given by \cite{LeCorre10} recently. In order to understand DNB, the heat transfer and phase change should be first investigated in boiling flow. \cite{Kurul90} proposed a wall heat partitioning model which explained the heat transfer and mass transfer in the region near wall surface. \cite{Hibiki02} proposed a model of interfacial area concentration to study the bubble size in the bulk, which could also another aspect to influence the heat transfer. Three dimensional (3D) modeling of subcooled boiling flows could benefit from the open source CFD (OpenFOAM) nowadays. In our current solver, several models about interfacial momentum transfer, interfacial area concentration transportation, and wall heat partitioning are implemented.

One of the important issues of the current and future sustainable energy systems is the efficiency and stability of heat removal due to natural or mixed convection, forced convection or boiling heat transfer. In some energy systems natural heat convection is envisaged during normal operation. This type of heat removal is very reliable since it doesn't depend on availability of external pumping resources, and coolant flow through the system is assured by the gravity force. The drawback of the natural circulation is its inherent instability and also relatively low heat transfer efficiency.
Thus, in many high heat flux technologies, such as e.g. nuclear reactors, the boiling heat transfer is preferred as the most efficient heat transfer mode. The design of high heat flux systems requires a thorough fluid flow and heat transfer analysis in complex geometries. Traditionally experimental methods have been used for these purposes in the past. The drawback of such methods is their large cost and time consumption, inherently related to all required experimental work. In addition, experimental methods are rather difficult to be used for a design optimization, where various geometry and/or operation condition variations are to be tested. For such purposes the most efficient design and optimization approach is based on computational tools, which are able to capture the geometry details and to include the governing phenomena. Currently the computational fluid dynamics (CFD) technology is widely used to design and to optimize heat transfer and fluid flow systems if single-phase flow conditions prevail. For two-phase flow applications, and in particular for boiling heat transfer conditions the CFD technology is still not mature enough. In particular, there is still lack of thoroughly validated and generally valid closure laws for subcooled and saturated nucleate flow boiling heat transfer, with a potential to be extended to predict the departure from nucleate boiling (DNB). The major aim of this paper is to contribute with new model development and validation in this particular area using open source CFD code OpenFOAM.
The first model suitable for CFD applications was developed by \cite{Kurul90}, who proposed a heat flux partitioning scheme to separately deal with vapor generation, sensible heat and quenching terms in the proximity of the heated wall. In the bulk bubbly flow, \cite{Hibiki02} proposed a two-equation model to predict the bubble size (and thus the interfacial area concentration) as a function of local flow conditions.

\section{Field equation in two-phase bubbly flow}
The present model includes mass, linear momentum and energy conservation equations for liquid and vapor phase. In addition, transport equations for the interfacial area concentration and for the turbulence are used to close the model. The details of the employed governing equations are given below.

\subsection{Phase continuity equation}

\begin{equation}\label{alphaeqn1}
\frac{\partial ( \alpha_k \rho_k)}{\partial t}+\nabla \cdot (\alpha_k  \rho_k  {\mathbf  U}_k )=\Gamma_{k}
\end{equation}
$\Gamma_k $ means the mass gained by phase $k$. ($k=l,v$)

\subsection{Linear momentum conservation equation}

\begin{align}\label{ueqn}
\frac{\partial ( \alpha_k {\rho_k} {{\mathbf U}_k}   )}{\partial t}+\nabla \cdot (\alpha_k {\rho_k} {{\mathbf U}_k}{{\mathbf U}_k})=-\alpha_k \nabla {p}+ \nabla \cdot \left[\alpha_k ({{\bm \tau}_k} + {\bm \tau}_k^{\rm t})\right]
+\alpha_k {\rho_k} \mathbf g + \Gamma_{k} {{ \mathbf U}_{ki}} + \mathbf{M}_{ki}
\end{align}

Here the interfacial velocity is modeled as
\begin{equation}
{ \mathbf U}_{ki}=
\begin{cases}
    { \mathbf U}_{l} & \text{if}\ \Gamma_{v}>0,\ \text {evaporation} \\
    { \mathbf U}_{v} & \text{if}\ \Gamma_{v}<0,\ \text {condensation}
\end{cases}
\end{equation}
using the upwind scheme.

According to the Boussinesq hypothesis, the turbulent stress strain relation is analogous to that of Newtonian fluids and consequently the effective stress appears as a function of fluid properties and velocity, which is used by \cite{Rusche} in OpenFOAM,

\begin{equation}
{\bm \tau}_k^{\rm eff}={{\bm \tau}_k} + {\bm \tau}_k^{\rm t}=\rho_k\nu_k^{\rm eff}\left( \nabla { \mathbf U}_{k} + \left( \nabla{ \mathbf U}_{k}\right)^T - \frac{2}{3}\mathbf I \nabla \cdot{ \mathbf U}_{k} \right)-\frac{2}{3}\mathbf I \rho_k k_k
\end{equation}
and,
\begin{equation}
\nu_k^{\rm eff}=\nu_k+\nu_k^{\rm t}
\end{equation}

\subsection{Enthalpy equation}

\begin{equation}\label{enthalpy}
\frac{\partial ( \alpha_k {\rho_k}{h_k})}{\partial t}+\nabla \cdot (\alpha_k {\rho_k} {h_k} {\mathbf U}_k )=-\nabla \cdot \left[ \alpha_k ({\mathbf q}_k^{\prime \prime}+{\mathbf q}_k^{\rm t}) \right]+\Gamma_{k} h_{ki}+a_i { {q}^{\prime\prime}_{ki}}+a_w { {q}^{\prime\prime}_{kw}}
\end{equation}
where $a_w$ refers to heated area per unit controlled volume of fluid between the wall and the liquid phase.

\cite{Kurul91} discussed the mass conservation and energy conservation at the interface and first proposed the corresponding equations in two-phase flow. Here we formulate the mass flux $\Gamma_l$ from phase $v$ to phase $l$ furthermore as,

\begin{equation}\label{gamal}
\Gamma_{l}=
\begin{cases}
    \displaystyle\frac{a_i q_{li}^{\prime\prime}+a_iq_{vi}^{\prime\prime}}{h_{v}-h_{l,\rm sat}} &
%\text{if}\ \Gamma_l>0, \ {\rm i.e.} \
%\text{if}\ a_i q_{li}^{\prime\prime}+a_iq_{vi}^{\prime\prime}>0,\
\text {condensation} \\
    \displaystyle\frac{a_i q_{li}^{\prime\prime}+a_iq_{vi}^{\prime\prime}}{h_{v,\rm sat}-h_{l}} &
%\text{if}\ \Gamma_l<0, \ {\rm i.e.} \
%\text{if}\ a_i q_{li}^{\prime\prime}+a_iq_{vi}^{\prime\prime}<0,\
\text {evaporation}
\end{cases}
\end{equation}
where the interfacial enthalpy $h_{ki}\:(k=l,v)$ is modeled with the upwind approximation. The modeling of interfacial heat transfer $a_i q_{li}^{\prime\prime}$ and $a_i q_{vi}^{\prime\prime}$ will be introduced in the following section.

Equation \ref{gamal} could be applied to the heat transfer in the bulk. For those cells which are adjacent to the wall directly, we have totally different heat transfer mechanism since there are interaction among the liquid, vapor and walls. Here we assume that only evaporation is allowed in those cells, which is consistent with the situation in boiling flows. In those cells, the total heat transfer per unit volume to phase $l$ is given as,
\begin{equation}
q^{\prime\prime\prime}_l =a_i q_{li}^{\prime\prime} -\Gamma_{vl}h_{l}+a_w { {q}^{\prime\prime}_{lw}}
\end{equation}
and the total heat transfer to phase $v$ as,
\begin{equation}
q^{\prime\prime\prime}_v =a_i q_{vi}^{\prime\prime} +\Gamma_{vl}h_{v}+a_w { {q}^{\prime\prime}_{vw}}
\end{equation}

The energy balance in those cells could be written as
\begin{equation}
%\Gamma_{vl} h_{v}-\Gamma_{vl} h_{l}+a_i { {q}^{\prime\prime}_{vi}}+a_i { {q}^{\prime\prime}_{li}}+a_w { {q}^{\prime\prime}_{lw}}-a_w{q}^{\prime\prime}_{w}=0
q^{\prime\prime\prime}_l+q^{\prime\prime\prime}_v=a_w{q}^{\prime\prime}_{w}
\end{equation}

Usually we make an assumption that in subcooled flow boiling, the temperature of the vapor phase is constant and equal to the saturation temperature. In addition, we neglect a direct heating of vapor from the wall, that is: $a_w { {q}^{\prime\prime}_{vw}}=0$. With these assumptions it is straightforward to calculate the heat flux to each phase in cells adjacent to the heated walls.

%And it is easy to calculate heat flux to each phase by given $\Gamma_{vl}$ in the cells adjacent to walls.

Using the Fourier's law of conduction for the liquid phase, the molecular heat flux in Eqn. \ref{enthalpy} can be written as,
\begin{equation}
{\mathbf q}_k^{\prime \prime}=-\frac{\lambda_l}{c_{pl}}\nabla h_l
\end{equation}
where $\lambda$ and $c_p$ are respectively the thermal conductivity and the specific heat.

The turbulent heat flux is found as follows,
\begin{equation}
{\mathbf q}_l^{\rm t}=-\frac{\lambda_l^{\rm t}}{c_{pl}}\nabla h_l
\end{equation}
where the turbulent thermal conductivity is given as,,
\begin{equation}
{\lambda_l^{\rm t}}=\frac{c_{pl}\rho_l\nu_l^{\rm t}}{ {\rm Pr}_l^{\rm t}}
\end{equation}
where Pr$_l^{\rm t}$ is the turbulent Prandtl number of phase $l$. A constant value of 0.9 has been chosen for Pr$_l^{\rm t}$ in the calculations presented in this paper.

In OpenFOAM, equation \ref{enthalpy} of liquid phase is reorganized into a phase intensive form,

\begin{align}\label{hl2}
&\frac{\partial{h_l}}{\partial t}+{\mathbf U}_l \cdot \nabla {h_l} - \nabla \cdot (\kappa_l^{\rm eff}\nabla h_l)-\kappa_l^{\rm eff} \frac{\nabla (\beta\rho_l)}{\beta\rho_l}\cdot \nabla h_l \cr
=&\begin{cases}
\displaystyle\frac{\Gamma_{lv} h_{l,\rm sat}- \Gamma_{lv} h_l + a_i {q}^{\prime\prime}_{li} }{\beta\rho_l}  &\text {bulk condensation} \\
\displaystyle\frac{a_i {q}^{\prime\prime}_{li} }{\beta\rho_l}  &\text {bulk evaporation} \\
\displaystyle\frac{a_i {q}^{\prime\prime}_{li} +a_w { {q}^{\prime\prime}_{lw}}}{\beta\rho_l} &\text {near wall cells}
\end{cases}
\end{align}
where,
\begin{equation}
\kappa_l^{\rm eff}= \frac{\lambda_l}{\rho_l c_{pl}}+\frac{\nu_l^{\rm t}}{{\rm Pr}_l^{\rm t}}
\end{equation}

The term $\displaystyle\frac{a_w { {q}^{\prime\prime}_{lw}}}{\beta\rho_l}$ on the right hand side (RHS) of Eqn. \ref{hl2} results from the thermal boundary condition at heated walls. Thus we treat this term by a gradient boundary condition in the energy transport equation.

In a similar manner, equation \ref{enthalpy} of the vapor phase is given as follows,
\begin{align}\label{hv}
&\frac{\partial{h_v}}{\partial t}+{\mathbf U}_v \cdot \nabla {h_v} - \nabla \cdot (\kappa_v^{\rm eff}\nabla h_v)-\kappa_v^{\rm eff} \frac{\nabla (\alpha\rho_v)}{\alpha\rho_v}\cdot \nabla h_v \cr
=&\begin{cases}
\displaystyle\frac{ a_i {q}^{\prime\prime}_{vi} }{\alpha\rho_v}  &\text {bulk condensation}\\
\displaystyle\frac{\Gamma_{vl} h_{v,\rm sat}- \Gamma_{vl} h_v + a_i {q}^{\prime\prime}_{vi} }{\alpha\rho_v}  &\text {bulk evaporation }\\
\displaystyle \frac{ a_i {q}^{\prime\prime}_{vi} }{\alpha\rho_v}  &\text {near wall cells}
\end{cases}
\end{align}
where,
\begin{equation}
\kappa_v^{\rm eff}= \frac{\lambda_v}{\rho_v c_{pv}}+\frac{\nu_v^{\rm t}}{{\rm Pr}_v^{\rm t}}
\end{equation}

\subsection {Interfacial area concentration transport equation}
The interfacial area concentration corresponds to the area of the gas bubbles per unit volume. For spherical bubbles,
\begin{equation}
a_i=\frac{6\alpha}{D_S}
\end{equation}
where $D_S$ is the bubble Sauter diameter, equal to the diameter of a sphere of an equivalent volume.

\cite{Hibiki02} modeled sink and source terms of the interfacial area concentration based on mechanisms of bubble-bubble and bubble-turbulent eddy random collisions, and they also introduced the effect by gas expansion,
\begin{equation}\label{ai}
\frac{\partial a_i}{\partial t}+\nabla \cdot (a_i {\mathbf U}_v )=\frac{2}{3}\frac{a_i}{\alpha}\left( \frac{\partial \alpha}{\partial t}+\nabla \cdot (\alpha {\mathbf U}_v)  \right)+ \Phi_{\rm BB}+ \Phi_{\rm BC}+\Phi_{\rm NUC}
\end{equation}
The first term on the RHS of Eqn. \ref{ai} refers to the contribution of phase change and expansion due to the pressure change. $\Phi_{\rm BB}$ and $\Phi_{\rm BC}$ represent the bubble number variations induced by the breakup and coalescence phenomena, respectively. In the \cite{Hibiki02} model, they are defined as,
\begin{equation}
\Phi_{\rm BC} = -\frac{1}{3\psi}\left(\frac{\alpha}{a_i}\right)^2\cdot \Gamma_C \frac{\alpha^2 \epsilon_l^{1/3}}{D_S^{11/3}(\alpha_{\max}-\alpha)}\exp \left( -K_C \frac{D_S^{11/3}\rho_l^{1/2}\epsilon_l^{1/3}}{\sigma^{1/2}} \right)
\end{equation}
with $\Gamma_C=0.0314$ and $K_C=1.29, \alpha_{\max}=0.74$, and
\begin{equation}
\Phi_{\rm BB} = \frac{1}{3\psi}\left(\frac{\alpha}{a_i}\right)^2\cdot \Gamma_B \frac{\alpha(1-\alpha) \epsilon_l^{1/3}}{D_S^{11/3}(\alpha_{\max}-\alpha)}\exp \left( -K_B \frac{\sigma} {D_S^{5/3}\rho_l\epsilon_l^{2/3}}\right)
\end{equation}
with $\Gamma_B=0.0209$ and $K_B=1.59$. Here $\psi=1/(36\pi)$ for spherical bubbles.

$\Phi_{\rm NUC}$ refers to an increase of interfacial area concentration by a bubble nucleation at the heated wall. \cite{Bae08} proposed the nucleation source term as,
\begin{equation}
\Phi_{\rm NUC} = \pi d_{lo}^2 \cdot {N^{\prime \prime} f  a_w}
\end{equation}
where $d_{lo}$ is the bubble lift-off diameter, $N^{\prime \prime}$ the active nucleation site density, and $f$ the bubble departure frequency.

\cite{Yao} proposed the breakup and coalescence term as,

\begin{equation}
\Phi_{\rm BC} = -\frac{1}{3\psi}\left(\frac{\alpha}{a_i}\right)^2\cdot K_{c1}\frac{\alpha^2 \epsilon_l^{1/3}}{D_S^{11/3}}
\frac{1}{1-(\alpha/\alpha_{\max})^{1/3}+K_{c2}\alpha \sqrt{{\rm We}/{{\rm We}_{cr}}}}
\exp\left( -K_{c3} \sqrt{\frac{\rm We}{{\rm We}_{cr}}}\right)
\end{equation}
where $K_{c1}$ = 2.86, $K_{c2}$ = 1.922, $K_{c3}$ = 1.017, ${\rm We}_{cr}$ = 1.24 and $\alpha_{\max}$ = 0.52.
\begin{equation}
\Phi_{\rm BB} = \frac{1}{3\psi}\left(\frac{\alpha}{a_i}\right)^2\cdot K_{b1}\frac{\alpha(1-\alpha) \epsilon_l^{1/3}}{D_S^{11/3}} \frac{1}{1+K_{b2}(1-\alpha)\sqrt{{\rm We}/{\rm We}_{cr}}}   \exp \left( -\frac{{\rm We}_{cr}}{{\rm We}} \right)
\end{equation}
where $K_{b1}=1.6$, and $K_{b2}=0.42$.

\cite{Lo} proposed a $S_\gamma$ model in which the breakup terms can be written down as,
\begin{equation}
\Phi_{\rm BB} = \pi\int_{D_{S_{cr}}}^\infty \frac{(2^{1/3}-1)D_S^2}{\tau_{br}}nP{\rm d}D_S
\end{equation}
Here $n=\displaystyle\frac{6\alpha}{\pi D_S^3}$ is the bubble number density. $P$ represents the log-normal distribution of bubble diameter,
\begin{equation}
P=\frac{1}{\sqrt{2\pi}D_S\hat\sigma} \exp \left( -\frac{({\rm ln}D_S -{\rm ln}\overline {D_S})^2}{2\hat\sigma^2} \right)
\end{equation}
where we use $\hat\sigma=0.5$ in the current solver. In the original paper, the breakup source term is modeled in two regimes: the viscous breakup regime and inertia breakup regime. And the overall source term should be summed up over the two regimes. However, since the mechanism is not well explained in the viscous regime, only the inertial breakup part is included in the current solver.

The Kolmogorov length scale $L_k$ is used to evaluate the regime that breakup takes place.
\begin{equation}
L_k=\left( \frac{\nu^3}{\epsilon} \right)^{1/4}
\end{equation}
Considering that only those bubbles of big size can break, the critical size in the inertia regime becomes as follows,
\begin{equation}
D_{Scr}=(1+C_\alpha\alpha)\left( \frac{\sigma {\rm We}_{cr}}{2\rho_l}  \right)^{3/5} \epsilon ^{-2/5}
\end{equation}

\begin{equation}
\tau_{br}= 2\pi k_{br}\sqrt{\frac{(3\rho_v+2\rho_l)D_S^3}{192\sigma}}
\end{equation}
with $C_{\alpha}=0$ and $k_{br}=0.2$.

The source term from bubble coalescence is modeled as,
\begin{equation}
\Phi_{\rm BC} = \pi(2^{1/3}-2)\left(\frac{6\alpha}{\pi}\right)^2k_{coll}U_r  P_{coal}D_S^{-2}
\end{equation}
where,
\begin{equation}
k_{coll}=\left(\frac{2\pi}{15}\right)^{1/2}
\end{equation}

\begin{equation}
U_r = (\epsilon D_S)^{1/3}
\end{equation}

\begin{equation}
P_{coal} = \frac{\Phi_{\rm max}}{\pi}\left( 1-\frac{k_{cl,2}^2 ({\rm We}-{\rm We}_0)^2}{16\Phi_{\rm max}^2}  \right)^{1/2}
\end{equation}

\begin{equation}
\Phi_{\rm max}=\frac{8h_0^2\rho_l\sigma}{{\rm We}_0\mu_v^2D_S}
\end{equation}
with the following coefficient:
$k_{cl,2}=12.7$, ${\rm We}_0=0.8{\rm We}_{cr}$ and $h_0=8.3h_{cr}$
\begin{equation}
h_{cr}=\left( \frac{A_HD_S}{24\pi\sigma} \right)^{1/3}
\end{equation}
where $A_H=5.0 \times 10^{-21}$ is the Hamaker constant.

%Recently, \cite{Park} formulated an analytical form of interfacial area concentration transport equation, in which the sink term due to condensation is considered. The model has not been included in our current solver since it is a one dimensional model.

%
%
\subsection{Turbulence modeling}
\subsubsection{Turbulence of liquid phase}

\cite{Rusche} proposed the standard $k-\epsilon$ model as follows,

\begin{equation}\label{tk0}
\frac{\partial (\beta \rho_l k_l)}{\partial t}+\nabla \cdot (\beta \rho_l \mathbf U_l k_l)
=\nabla \cdot \left[ \beta \left( \frac{\mu_l^{\rm eff}}{\sigma_k} \right)\nabla k_l \right]+\beta G-\beta \rho_l \epsilon_l
\end{equation}

\begin{equation}\label{teps0}
\frac{\partial (\beta \rho_l \epsilon_l)}{\partial t}+\nabla \cdot (\beta \rho_l \mathbf U_l \epsilon_l)
=\nabla \cdot \left[ \beta \left( \frac{\mu_l^{\rm eff}}{\sigma_\epsilon}\right)\nabla \epsilon_l \right]+\frac{\beta \epsilon_l}{k_l}(C_{\epsilon 1}G-C_{\epsilon 2}\rho_l \epsilon_l)
\end{equation}

Here $G$ stands for the production of turbulent kinetic energy and is defined as,
\begin{equation}
G=2\mu_l^{\rm t}\left(\nabla \mathbf U_l \cdot {\rm dev}(\nabla \mathbf U_l + (\nabla \mathbf U_l)^T)\right)
\end{equation}

%Liquid turbulence is further modeled with additional source terms to incorporate the effects of the dispersed phase on the turbulence by \cite{Yao},

In the above model, no effect of the dispersed phase on the turbulence in the continuous phase is taken into account. This deficiency is removed in the model proposed by \cite{Yao}, where an additional source term, representing the above-mentioned effect, is included,

\begin{align}\label{tk}
\frac{\partial (\beta \rho_l k_l)}{\partial t}+\nabla \cdot (\beta \rho_l \mathbf U_l k_l)
=&\nabla \cdot \left[ \beta \left( \frac{\mu_l^{\rm t}}{\sigma_k} \right)\nabla k_l \right]
-\beta \rho_l \epsilon_l + \beta {\bm \tau}_l : \nabla {\mathbf U}_l\cr
&- (\mathbf M_v^d + \mathbf M_v^{vm})\cdot(\mathbf U_v - \mathbf U_l)-\sigma (\Phi_{\rm BC}+\Phi_{\rm BB})+k_{li}\Gamma_l
\end{align}

\begin{align}\label{teps}
\frac{\partial (\beta \rho_l \epsilon_l)}{\partial t}+\nabla \cdot (\beta \rho_l \mathbf U_l \epsilon_l)
=&\nabla \cdot \left[ \beta \left( \frac{\mu_l^{\rm t}}{\sigma_\epsilon}\right)\nabla \epsilon_l \right]-C_{\epsilon 2}\beta\rho_l\frac{ \epsilon_l^2}{k_l} + C_{\epsilon 1}   \beta \frac{ \epsilon_l}{k_l}  {\bm \tau}_l : \nabla {\mathbf U}_l-\frac{2}{3}  \beta \rho_l {\epsilon_l} \nabla \cdot {\mathbf U}_l\cr
&- C_{\epsilon 3}(\mathbf M_v^d + \mathbf M_v^{vm})\cdot(\mathbf U_v - \mathbf U_l) \left ( \frac{\epsilon_l}{D_S^2} \right )^{1/3}+\epsilon_{li}\Gamma_l
\end{align}

The liquid Reynolds stress tensor is modeled as,
\begin{equation}
{\bm \tau}_l = \rho_l \nu_l^{\rm t} \left(   \nabla {\mathbf U}_l  + (\nabla {\mathbf U}_l)^T \right)-\frac{2}{3}\rho_l \left( k_l+\nu_l^{\rm t} \nabla \cdot {\mathbf U}_l \right){\mathbf I}
\end{equation}

The turbulent viscosity of liquid phase is given by \cite{Sato} as,
\begin{equation}
\nu_l^{\rm t} = C_\mu \frac{k^2}{\epsilon}+\frac{1}{2}C_{\mu b}D_S \alpha |\mathbf U_v - \mathbf U_l|
\end{equation}

The coefficients used in this work are $\sigma_k=1.0$, $\sigma_\epsilon=1.3$, $C_{\epsilon 1}=1.44$, $C_{\epsilon 2}=1.92$, $C_{\epsilon 3}=0.6$, $C_\mu=0.09$ and $C_{\mu b}=1.2$.

\subsubsection{Turbulence of vapor phase}

The turbulence of vapor phase is assumed to be dependent on that of the liquid phase. To this end, a turbulence response coefficient $C_t$, defined as the ratio of the root mean square values of dispersed phase velocity, is introduced. In this approach, the effective viscosity of the vapor phase is expressed as
\begin{equation}
\nu_v^{\rm eff}=\nu_v+C_{t}^2\nu_l^{\rm t}
\end{equation}
%$C_t$ could be modeled as function of local parameters such as void fraction. Here it is set to zero.
In a more elaborated model, $C_t$ could be calculated as a function of local parameters, such as e.g. void fraction. However, in the present approach the influence of the liquid phase is neglected and $C_t$ is set equal to zero.
\section{Interfacial momentum transfer closure laws}
The interfacial forces acting on a bubble are caused by the liquid which surrounds it. Ignoring the effect of the change of the mean curvature on the mixture momentum source, we have,
\begin{equation}
\mathbf M_v +\mathbf M_l=0
\end{equation}

The closure relationships for the interfacial forces are expressed in terms of the following non-dimensional numbers,

Eotvos number,
\begin{equation}
{\rm Eo}=\frac{(\rho_l-\rho_v)gD_S^2}{\sigma}
\end{equation}

Reynolds number,
\begin{equation}
{\rm Re}_b = \frac{|\mathbf U_v -\mathbf U_l|D_S}{\nu_l}
\end{equation}

\begin{equation}
{\rm Re}_{bm} = \frac{\rho_l|\mathbf U_v -\mathbf U_l|D_S}{\mu_m}
\end{equation}
Here,
\begin{equation}
\mu_m = \mu_l \left(  1-\frac{\alpha}{\alpha_{\rm max}}  \right)^{-2.5\alpha_{\rm max}\mu^\ast}
\end{equation}
\begin{equation}
\mu^\ast=\frac{\mu_v+0.4\mu_l}{\mu_v+\mu_l}
\end{equation}

The interfacial momentum transfer terms include different kinds of forces, each of them representing a separate physical phenomenon, including the drag force, the lift force, the wall lubrication force, the turbulent dispersion force and the virtual mass force, which constitute the total interfacial force as follows,
\begin{equation}
\mathbf M_v = \mathbf M_v^d+\mathbf M_v^l+\mathbf M_v^{wl}+\mathbf M_v^{td}+\mathbf M_v^{vm}
\end{equation}

\subsection{Drag force}
This force represents a resistance of the relative motion between two phases.
\begin{equation}
\mathbf M_v^d = -\frac{3}{4}\frac{C_{ds}}{D_S}\rho_l \alpha |\mathbf U_v -\mathbf U_l|(\mathbf U_v -\mathbf U_l)
\end{equation}
The following two models for the drag force coefficient are included in the current solver:

\cite{Schiller35},
    \begin{equation}
    C_{ds}=\max \left(\frac{24}{{\rm Re}_b}(1+0.15{\rm Re}_b^{0.687}), 0.44 \right)
    \end{equation}

\cite{Ishii79},
    \begin{equation}
    C_{ds}=\max \left (\frac{24}{{\rm Re}_{bm}}(1+0.15{\rm Re}_{bm}^{0.687}), 0.44 \right)
    \end{equation}

\subsection{Lift force}
When a particle travels through the fluid with a non-uniform lateral velocity field, a lateral force will be acting between the fluid and the particle,
\begin{equation}
\mathbf M_v^l = C_l \rho_l \alpha (\mathbf U_v -\mathbf U_l)\times \nabla \times \mathbf U_l
\end{equation}

%The \cite{Tomiyama98} model is currently included in the solver,
In the present model the lift coefficient $C_l$ is calculated from the \cite{Tomiyama98} model,
\begin{equation}
C_l =
\begin{cases}
   {\rm min}(0.288{\rm tanh}(0.121{\rm Re}_b),f({\rm Eo}_d)) &  {{\rm Eo}_d < 4} \\
   f({\rm Eo}_d)&  {4<{\rm Eo}_d < 10} \\
   -0.27& {\rm Eo}_d > 10
\end{cases}
\end{equation}
\begin{equation}
f({\rm Eo}_d)=0.001509{\rm Eo}_d^3 -0.0159{\rm Eo}_d^2 -0.0204{\rm Eo}_d +0.474
\end{equation}
Here,
\begin{equation}
{\rm Eo}_d= \frac{(\rho_l-\rho_v)gd_h^2}{\sigma}
\end{equation}
\begin{equation}
d_h = D_S (1+0.163{\rm Eo}^{0.757})^{1/3}
\end{equation}

It should be noted that the force is turned off in the cells adjacent to walls in order to avoid unexpected fluctuation of void fraction in those cells in numerical simulation.

\subsection{Wall lubrication force}
This force was first proposed by \cite{Antal91} in order to explain the near wall void fraction features.
\begin{equation}
\mathbf M_v^{wl} =C_{w} \rho_l \alpha  |\mathbf U_r -(\mathbf U_r \cdot \mathbf  n_w )\mathbf  n_w |^2 \mathbf n_w
\end{equation}

The following two models for the wall lubrication force coefficient are included in the current solver:

\cite{Tomiyama98},
\begin{equation}
C_{w} = \frac{1}{2}C_{wl}D_S \left(  \frac{1}{y_w^2} - \frac{1}{(D_{pipe}-y_w)^2}     \right)
\end{equation}

\begin{equation}
C_{wl} =
\begin{cases}
   0.47&  {{\rm Eo} < 1} \\
   \exp (-0.933{\rm Eo}+0.179)&  {1<{\rm Eo}< 5} \\
   0.00599{\rm Eo}-0.0187& {5<{\rm Eo}< 33} \\
   0.179& {{\rm Eo} > 33}
\end{cases}
\end{equation}

\cite{Frank05},
\begin{equation}
C_{w} = C_{wl} \max \left(  0, \frac{1}{C_{wd}}\frac{1-y_w/C_{wc}D_S}{y_w(y_w/C_{wc}D_S)^{p-1}}     \right)
\end{equation}
It is suggested that $C_{wc}=10.0$, $C_{wd}=6.8$ and $p=1.7$.

\subsection{Turbulent dispersion force}
The turbulent dispersion force accounts for the turbulent fluctuations of the liquid phase and the effects, which the fluctuations have on the distribution of the gas phase. The following models are currently included in the solver:

\cite{Gosman92},
\begin{equation}
\mathbf M_v^{td} =-C_{d} \frac{3}{4} \frac{\rho_l}{D_S} \frac{\nu_l^t}{\sigma_\alpha}  |\mathbf U_r  | \nabla \alpha
\end{equation}

\cite{Bertodano92},
\begin{equation}
\mathbf M_v^{td} =-C_{td} \rho_l k_l \nabla \alpha
\end{equation}

\subsection{Virtual mass force}
\begin{equation}
\mathbf M_v^{vm} =-C_{vm} \rho_l \left( \frac{D\mathbf U_v }{Dt}-\frac{D\mathbf U_l }{Dt} \right)
\end{equation}
Currently it is assumed that $C_{vm}=0.5$.

\section{Liquid-vapor interfacial heat transfer closure laws}

\cite{Yao} proposed the following model for the liquid phase interfacial heat transfer,
\begin{equation}
a_i {q}^{\prime\prime}_{li} =
\begin{cases}
     c_{li} a_i (h_{l,\rm sat}-h_l) & \text{bulk} \\
    \text{not specified} & \text{near wall cells}
\end{cases}
\end{equation}
and,
\begin{equation}\label{cli}
 c_{li} = \frac{\lambda_l}{c_{pl}D_S}\rm Nu
\end{equation}

The Nusselt number is
\begin{equation}\label{Nu}
{\rm Nu}=
\begin{cases}
    2+0.6{\rm Re}^{0.5}{\rm Pr}^{0.33} & \text{if}\ {\rm Ja}<0,\ \text {condensation} \\
    \max(\rm Nu_1, Nu_2, Nu_3) & \text{if}\ {\rm Ja}>0,\ \text {evaporation}
\end{cases}
\end{equation}
where,
\begin{equation}
{\rm Ja}= \frac{\rho_lc_{pl}(T_l-T_{\rm sat})}{\rho_vh_{fg}},\:{\rm Re}=\frac{D_SU_r}{\nu_l},\:{\rm Pe}=\frac{D_SU_r}{\kappa_l}
\end{equation}
\begin{equation}
{\rm Nu}_1= \sqrt{\frac{4\rm Pe}{\pi}},\:{\rm Nu}_2= \frac{12}{\pi}{\rm Ja},\:{\rm Nu}_3= 2
\end{equation}

The interface to vapor heat transfer is expressed in the following manner,
%\begin{align}
%a_i {q}^{\prime\prime}_{vi}  &= c_{vi} (h_{v,\rm sat}-h_v) \cr
% &=\frac{\alpha\rho_v}{\delta t}(h_{v,\rm sat}-h_v)
%\end{align}

\begin{equation}
a_i {q}^{\prime\prime}_{vi}  = c_{vi} (h_{v,\rm sat}-h_v)
\end{equation}

\begin{equation}\label{cvi}
c_{vi} =\frac{\alpha\rho_v}{\delta t}
\end{equation}
where $\delta t$ is numerical time step. The above equations make sure that the vapor temperature is very close to the saturation temperature.
\section{Subcooled nucleate boiling model}
The wall heat transfer model for subcooled boiling flow was first proposed by \cite{Kurul90}, who partitioned the wall heat flux into three components: single phase convection, transient conduction as well as evaporation. The heat transfer coefficient for each process is correlated against experiment respectively. More recent work is done by \cite{Steiner05} and they believe that the total heat flux is assumed to be additively composed of a forced convective and a nucleate boiling component.

\subsection{Single phase convective heat transfer}
The single phase forced convection heat flux outside the influence area is calculated by \cite{Kurul90} as,
\begin{equation}\label{q1f}
q^{\prime\prime}_{c}=h_{fc}A_{1 \Phi}(T_w-T_l)
\end{equation}
where $h_{fc}$ is the single phase liquid heat transfer coefficient, $A_{1 \Phi}$ is the area fraction dominated by single phase convection, $T_w$ is wall temperature and $T_l$ is the subcooled liquid temperature.

The single phase forced convective heat transfer coefficient $h_{fc}$ is modeled as,
\begin{equation}\label{hfc}
h_{fc}=\rho_l c_{pl} \frac{u_{\tau}}{T^+}
\end{equation}
where the dimensionless temperature is modeled by \cite{Kader},
\begin{equation}
T^+ =\Pr y^+ \exp(-\eta)+(2.12 \ln y^+ + \beta_t) \exp(-1/\eta)
\end{equation}
and,
\begin{equation}
\beta_t=(3.85{\rm Pr}^{1/3}-1.3)^2+2.12\ln \Pr
\end{equation}

\begin{equation}
\eta=\frac{0.01(\Pr y^+)^4}{1+5{\rm Pr}^3y^+}
\end{equation}

\begin{equation}
y^+=\frac{\rho_lu_{\tau}|\mathbf U_l|}{\mu_l}
\end{equation}

The friction velocity is coupled with $k-\epsilon$ model,
\begin{equation}
u_{\tau}=C_\mu^{0.25}\sqrt{k};
\end{equation}

%%%%%%%%%%%%%%%%%%%%%%%%%%%%%%%%%%%%%%%%%%%%%%%%%%%%%%%%%%%%%%%%%%%%%%%%%%%%%%%%%%%%%%%%

\subsection{Quenching heat transfer}
The quenching (or transient conduction) heat flux is modeled as,
\begin{equation}\label{qq}
q^{\prime\prime}_{q}=h_q A_{b}(T_w-T_l)
\end{equation}
where $A_{b}$ represents the bubble influenced area fraction. According to \cite{Kurul90}, the bubble influenced area is determined by
\begin{equation}\label{area}
A_{b}=\min\left[ 1, N^{\prime\prime} K \left( \frac{\pi d^2_{lo}}{4}  \right) \right]
\end{equation}
%where $N^{\prime\prime}$ is the active nucleate site density and $d_{lo}$ is the bubbles lift off diameter.
Here $K$ determines the size of the bubble influence area around the nucleation site on the surface. $K=4$ is recommended by \cite{Victor}.

The quenching heat transfer coefficient is given by \cite{Victor},
\begin{equation}\label{hq}
h_q =2\frac{\lambda_l}{\sqrt{\pi \kappa_l t}}
\end{equation}
where $t=0.8/f$ represents the life span that the quenching heat flux experiences.

%%%%%%%%%%%%%%%%%%%%%%%%%%%%%%%%%%%%%%%%%%%%%%%%%%%%%%%%%%%%%%%%%%%%%%%%%%%%%%%%%%%%%%%%
\subsection{Evaporation heat transfer}
The evaporation rate is calculated as,
\begin{equation}\label{qe}
\Gamma_{vl}=\frac{\pi}{6}d_{lo}^2\rho_vfN^{\prime\prime}a_w
\end{equation}

%%%%%%%%%%%%%%%%%%%%%%%%%%%%%%%%%%%%%%%%%%%%%%%%%%%%%%%%%%%%%%%%%%%%%%%%%%%%%%%%%%%%%%%%
\subsection{Bubble detachment size}
%\citep{Unal, Situ05, Krepper11}
There are quite a few models to calculate the lift-off diameter \& departure diameter. \cite{Unal} made a correlation of bubble detachment diameter which is validated with pressure from 0.1 to 17.7 Mpa, heat flux from 0.47 to 10.64 MW/m$^2$, inlet velocity from 0.08 to 9.15 m/s, inlet subcooling from 3.0 to 86 K. \cite{Situ05} developed a bubble lift-off model based on force analysis. Their test runs were performed at 1 bar, and the model was validated with heat flux from 60.7 to 206 kW/m$^2$, inlet velocity from 0.487 to 0.939 m/s, inlet subcooling from 1.5 to 20 K. \cite{Krepper11} developed a correlation against the experimental data directly,
\begin{equation}
d_{lo}=d_{\rm ref}\exp\left(  -\frac{T_{\rm sat}-T_l}{\Delta T_{\rm refd}}  \right)
\end{equation}
where the reference value could be found at \cite{Krepper11} for certain experiment.

%%%%%%%%%%%%%%%%%%%%%%%%%%%%%%%%%%%%%%%%%%%%%%%%%%%%%%%%%%%%%%%%%%%%%%%%%%%%%%%%%%%%%%%%

\subsection{Bubble detachment frequency}
A simple estimation of the bubble departure frequency as the terminal rise velocity over the departure size is used here,

\cite{Ceumern},
\begin{equation}
f=\sqrt{\frac{4}{3}\frac{(\rho_l-\rho_v)g}{\rho_l d_{lo}}}
\end{equation}

%%%%%%%%%%%%%%%%%%%%%%%%%%%%%%%%%%%%%%%%%%%%%%%%%%%%%%%%%%%%%%%%%%%%%%%%%%%%%%%%%%%%%%%%

\subsection{Active nucleation site density}
A few models have been implement in the current solver \citep{Lemmert,Hibiki03,Krepper07,Krepper11}. Here the \cite{Krepper11} model is used for the validation.

\begin{equation}
N^{\prime\prime}=N_{\rm ref}\left( \frac{T_w-T_l}{\Delta T_{\rm refN}} \right)^p
\end{equation}
The reference value can be found in \cite{Krepper11}.

%%%%%%%%%%%%%%%%%%%%%%%%%%%%%%%%%%%%%%%%%%%%%%%%%%%%%%%%%%%%%%%%%%%%%%%%%%%%%%%%%%%%%%%%

\subsection{Liquid bulk temperature}
Another issue arises from the bulk liquid temperature. Here we used
\begin{equation}
T_{bulk}=T_w-\frac{T^{+}_{y^{+}_{bulk}}}{T^{+}_{y^{+}_{cell}}}(T_w-T_{cell})
\end{equation}
which is already implemented in ANSYS CFX5. The bulk temperature is obtained by setting $y^{+}_{bulk}=250$. Here the subscript {\it cell} refers to the cells adjacent to walls.

\section{Test cases}
%The Bartolomej experiment and DEBORA experiment are selected as the test case of the solver. Since it is difficult to find the original paper, the description of experiment conditions is found elsewhere \citep{KurulPhD,Krepper07,Yao, Krepper11}. The test section in Bartolomej experiment is a vertical pipe with 2 m in length and internal diameter 15.4 mm, while in DEBORA experiment it is a pipe with 3.5 m in length and internal diameter 19.2 mm. In Bartolomej experiment, the liquid temperature as well as void fraction were measured along the axis, while in DEBORA experiment, all the measurement were taken at the end of test section.
Two data sets were considered in calculations: the void fraction measurements performed by Bartolomej for subcooled boiling heat transfer to water under 45 bar pressure \citep{KurulPhD,Krepper07} and subcooled boiling heat transfer to refrigerant R-12 performed in the DEBORA experiment \citep{Yao, Krepper11}.

The experiment conditions used as test case are listed in Table \ref{expcondition}.

\begin{table}[h]
\caption{Selected test cases and their experiment conditions}
\begin{tabular}{cccccc}
\hline
  Case & Working fluid & Pressure  & Mass flow rate  & $q_w^{\prime\prime}$   & $T_{in}$  \\
   &  &  (bar) &  (kg/m$^2$/s) &   (kW/m$^2$) &  ($^\circ$C) \\
\hline
  Bart& water & 45 & 900 & 570 & 199.24  \\
  DEB5& R-12 & 26.15 & 1986 & 73.89 & 68.52  \\
  DEB6& R-12 & 26.15 & 1984.9 & 73.89 & 70.53  \\
\hline
\end{tabular}
\label{expcondition}
\end{table}

The tests were simulated in a quasi-two-dimensional cylindrical geometry, with 100 meshes in the axial direction and 20 meshes in the radial direction. The center of the grid cell adjacent to the wall has a non-dimensional coordinate of $y^+ = 60$ in Bartolomej test and $y^+ = 100$ in DEBORA test, approximately.  Grid refinement study performed by \cite{Krepper11} for the DEBORA experiment indicates that these values of $y^+$ provide grid-independent solutions. The boundary condition for liquid enthalpy adopted the {\it fixedGradient} type in order to account for the applied wall heat flux into liquid (see in Eqn. \ref{hl2}), as,
\begin{equation}
\nabla_f^\perp h_l = \frac{a_i {q}^{\prime\prime}_{li} +a_w { {q}^{\prime\prime}_{lw}}}{\beta\rho_l\kappa_l^{\rm eff}}
\end{equation}

The mass conservation and energy conservation over the whole pipe are carefully checked in the steady state. A typical error is
${\Delta G}/{G_{in}}=0.048\%$ and ${\Delta q^{\prime\prime}}/{q_{w}^{\prime\prime}}=1.6\%$.

In this test case, the following interfacial models are selected:
\begin{enumerate}
  \item Drag force: \cite{Ishii79}
  \item Lift force: \cite{Tomiyama98}
  \item Wall lubrication force: \cite{Tomiyama98}
  \item Turbulent dispersion force: \cite{Bertodano92}
\end{enumerate}

\begin{figure}
\subfloat[]
{
    \vspace*{0cm}\hspace*{-1cm}
    \epsfxsize=200pt\epsfbox{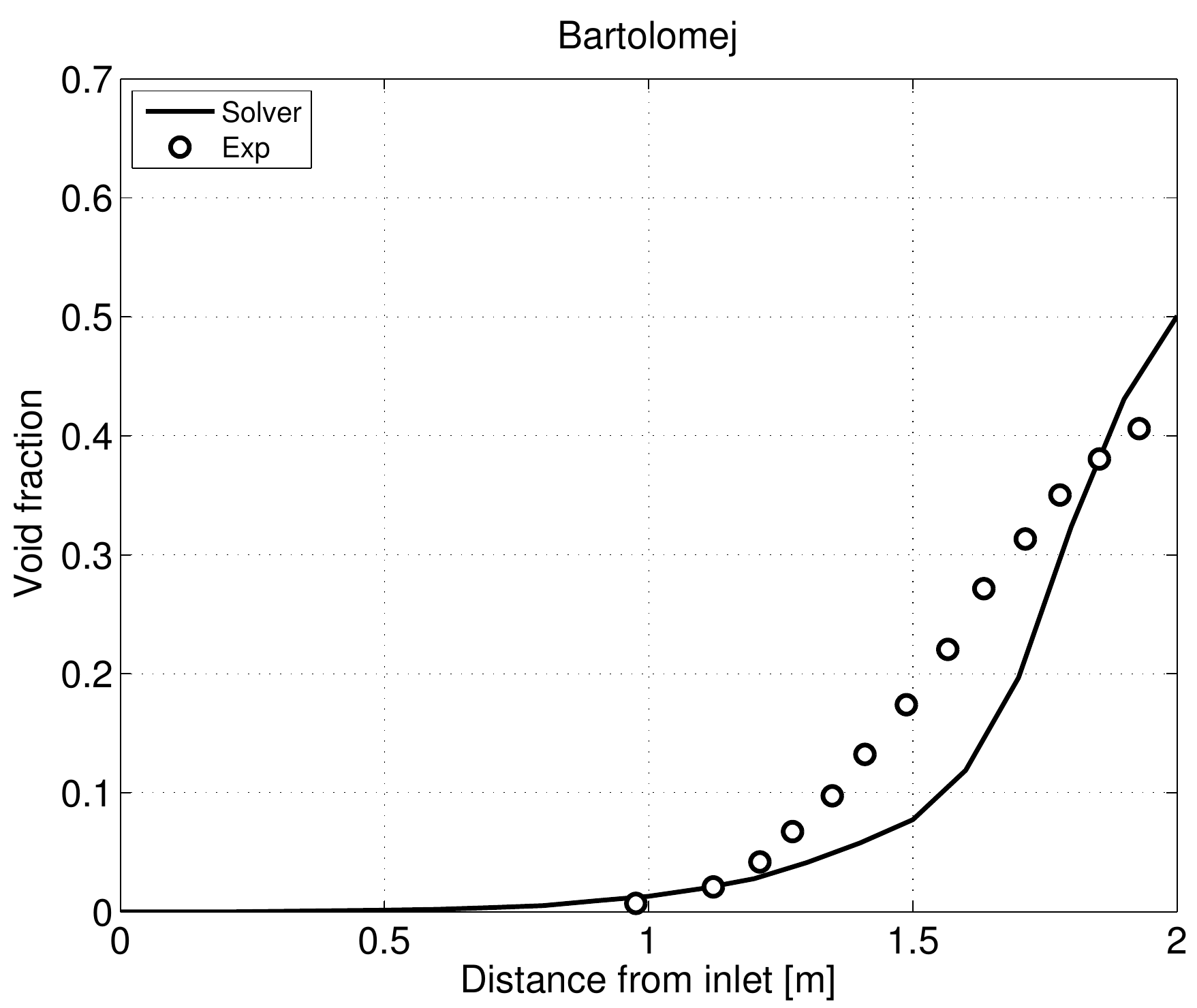}
    \vspace*{0cm}\hspace*{0cm}\label{fig1:a}
}
\subfloat[]
{
    \vspace*{0cm}\hspace*{0cm}
    \epsfxsize=205pt\epsfbox{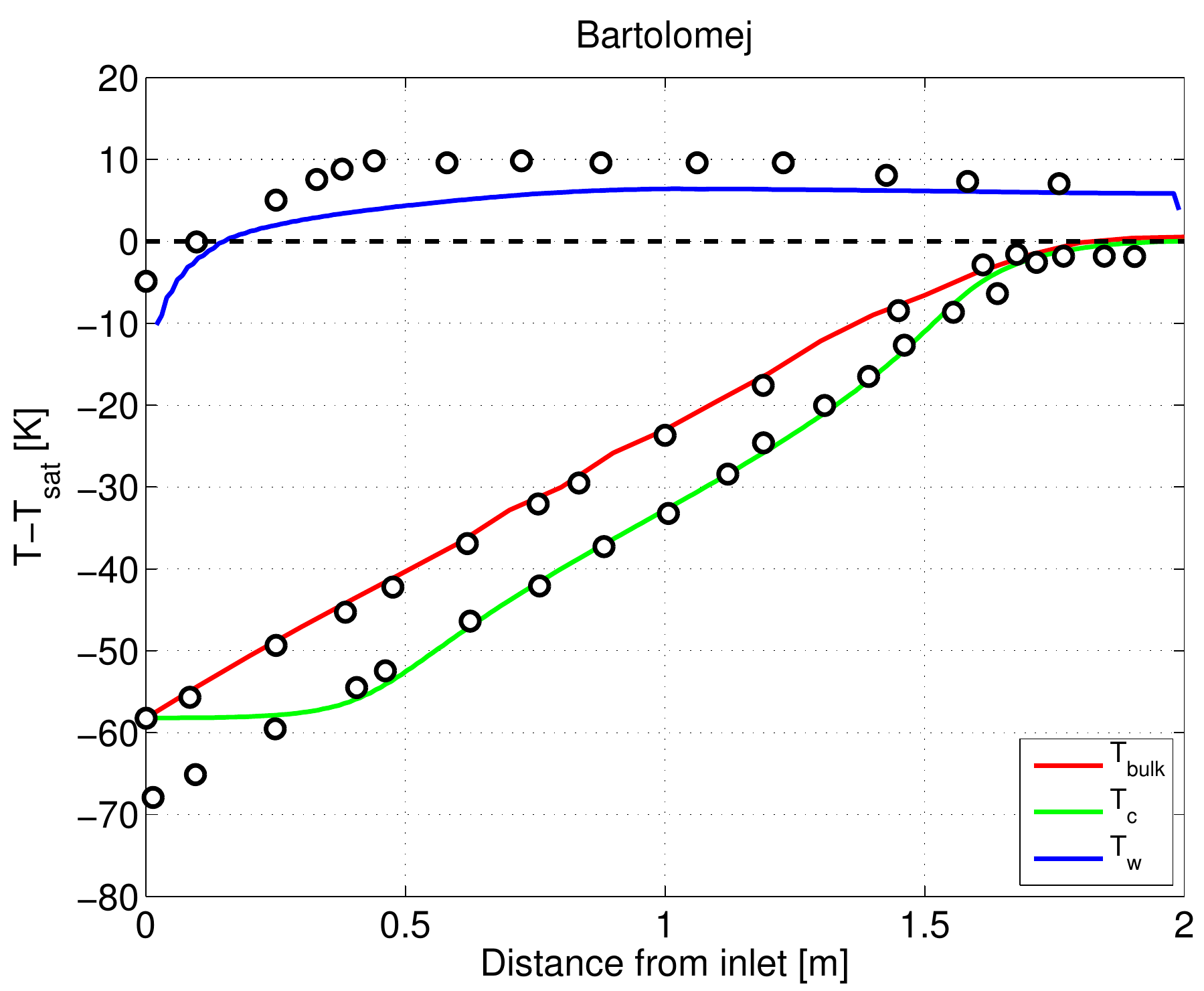}
    \vspace*{0cm}\hspace*{0cm}\label{fig1:b}
}
\caption{Axial steady state distribution of void fraction in Bartolomej experiment}
\label{fig1}
\end{figure}

Figure \ref{fig1} shows the comparison between the experimental and calculation results for the Bartolomej experiment, using the \cite{Yao} models with $C_{td}=2.5$. Since we used a uniformly distributed temperature profile as the inlet boundary condition for the energy conservation equation, there is a discrepancy between the predicted and measured temperature in the region nearby, as shown in Fig. \ref{fig1:b}. However, the temperature of the bulk and at the centerline could be well predicted after the flow becomes fully developed. The averaged void fraction is somehow underestimated, which may be due to several reasons. Firstly, we used a two-equation interfacial area concentration model in which the condensation rate could be overestimated due to underestimated bubble size. Unfortunately, the measurement of the bubble size is not available in the Bartolomej experiment, rending it difficult to evaluate the prediction of the bubble size. Secondly, the underestimation of void fraction could be also related to the modeling of interfacial forces, for example, turbulent dispersion force. If we have a large turbulent dispersion force that drives bubbles towards the cold bulk, the condensation could also be overestimated and results in a rather low void fraction. Thirdly, the observed discrepancy could also result from the underestimation of evaporation rate, which depends on the wall heat partitioning model.

\begin{figure}
\subfloat[]
{
    \vspace*{0cm}\hspace*{-1cm}
    \epsfxsize=200pt\epsfbox{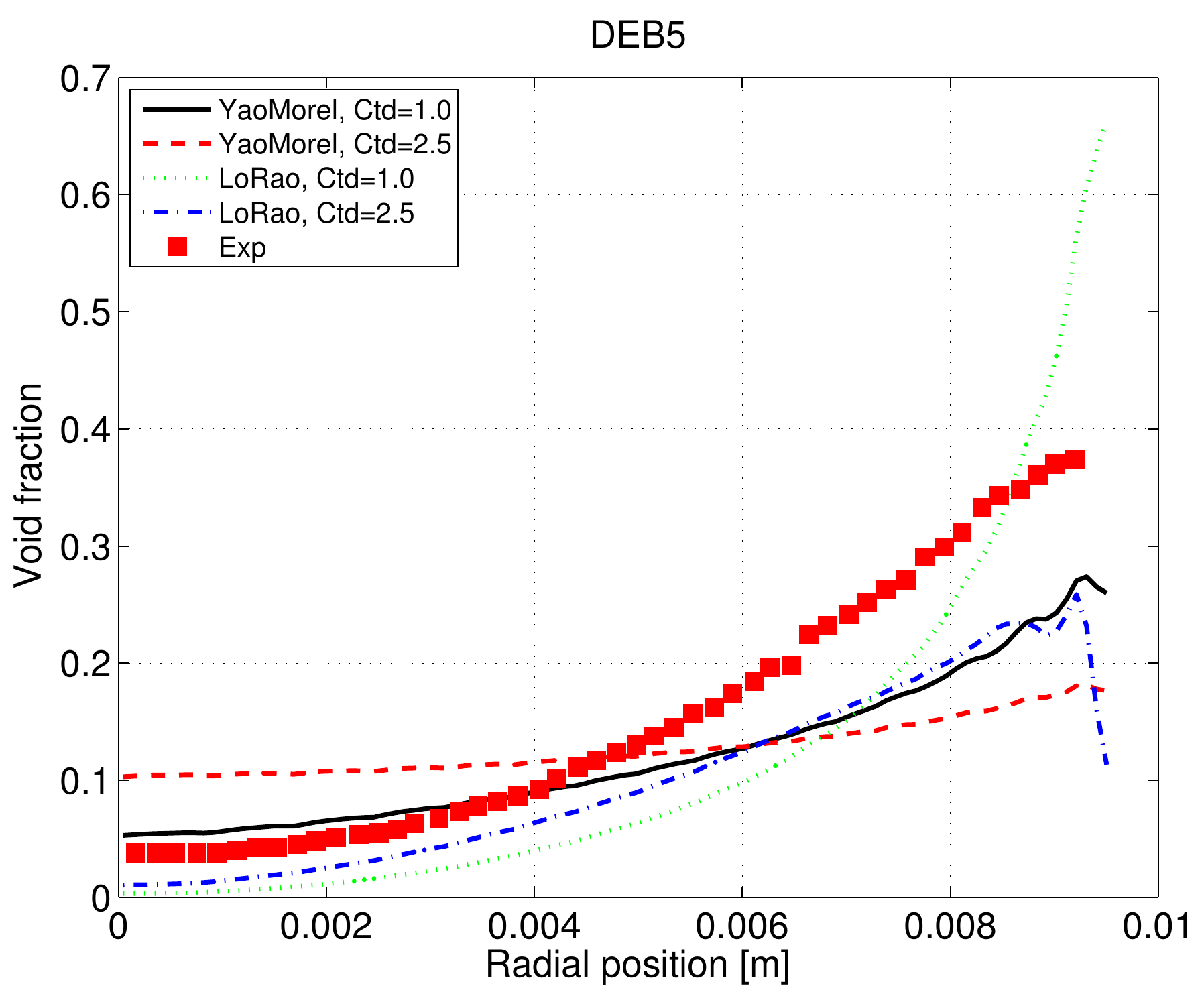}
    \vspace*{0cm}\hspace*{0cm}\label{fig2:a}
}
\subfloat[]
{
    \vspace*{0cm}\hspace*{0cm}
    \epsfxsize=200pt\epsfbox{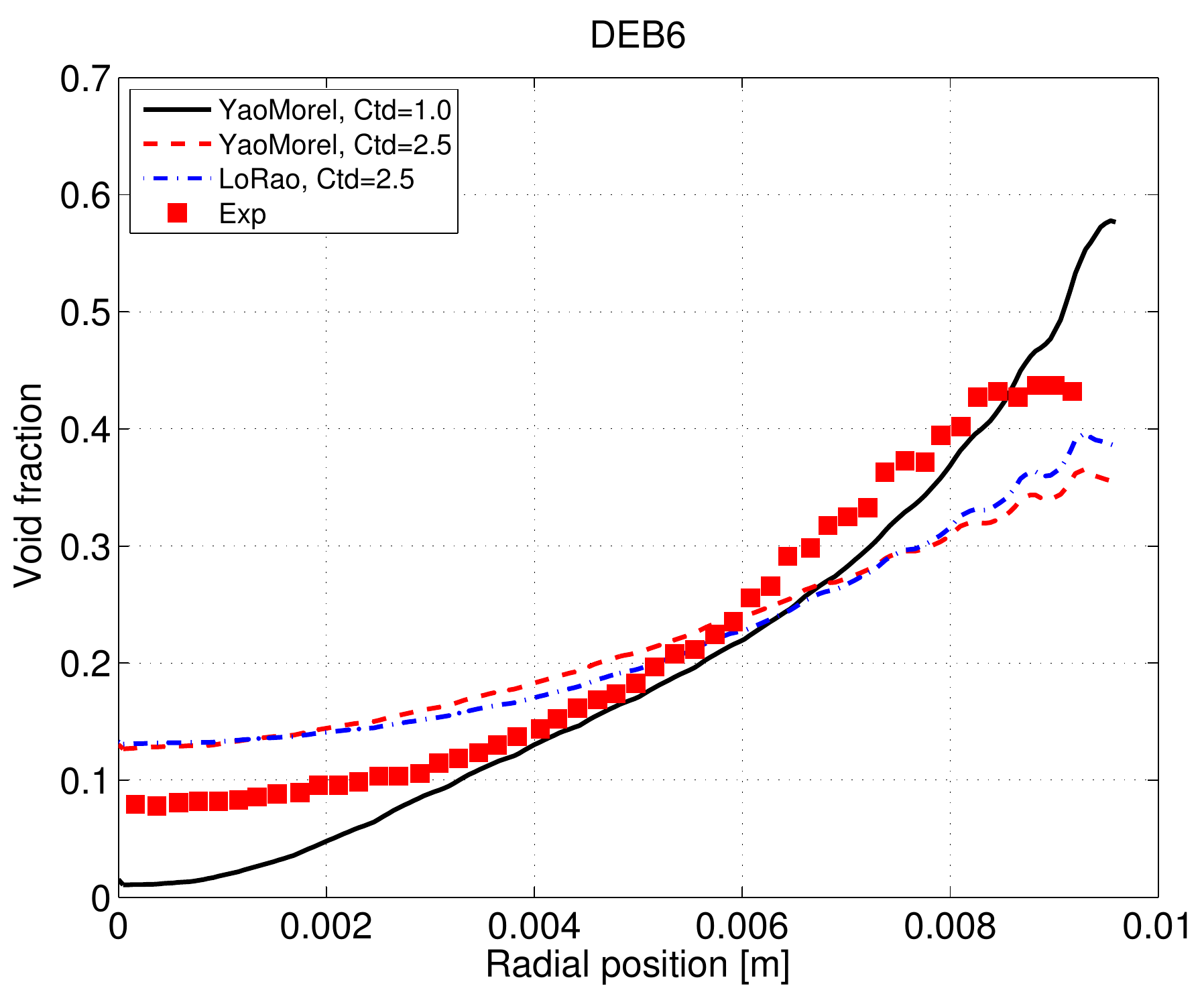}
    \vspace*{0cm}\hspace*{0cm}\label{fig2:b}
}
\caption{Comparison between the DEBORA experiment and calculation results: Radial void fraction}
\label{fig2}
\end{figure}

\begin{figure}
\subfloat[]
{
    \vspace*{0cm}\hspace*{-1cm}
    \epsfxsize=200pt\epsfbox{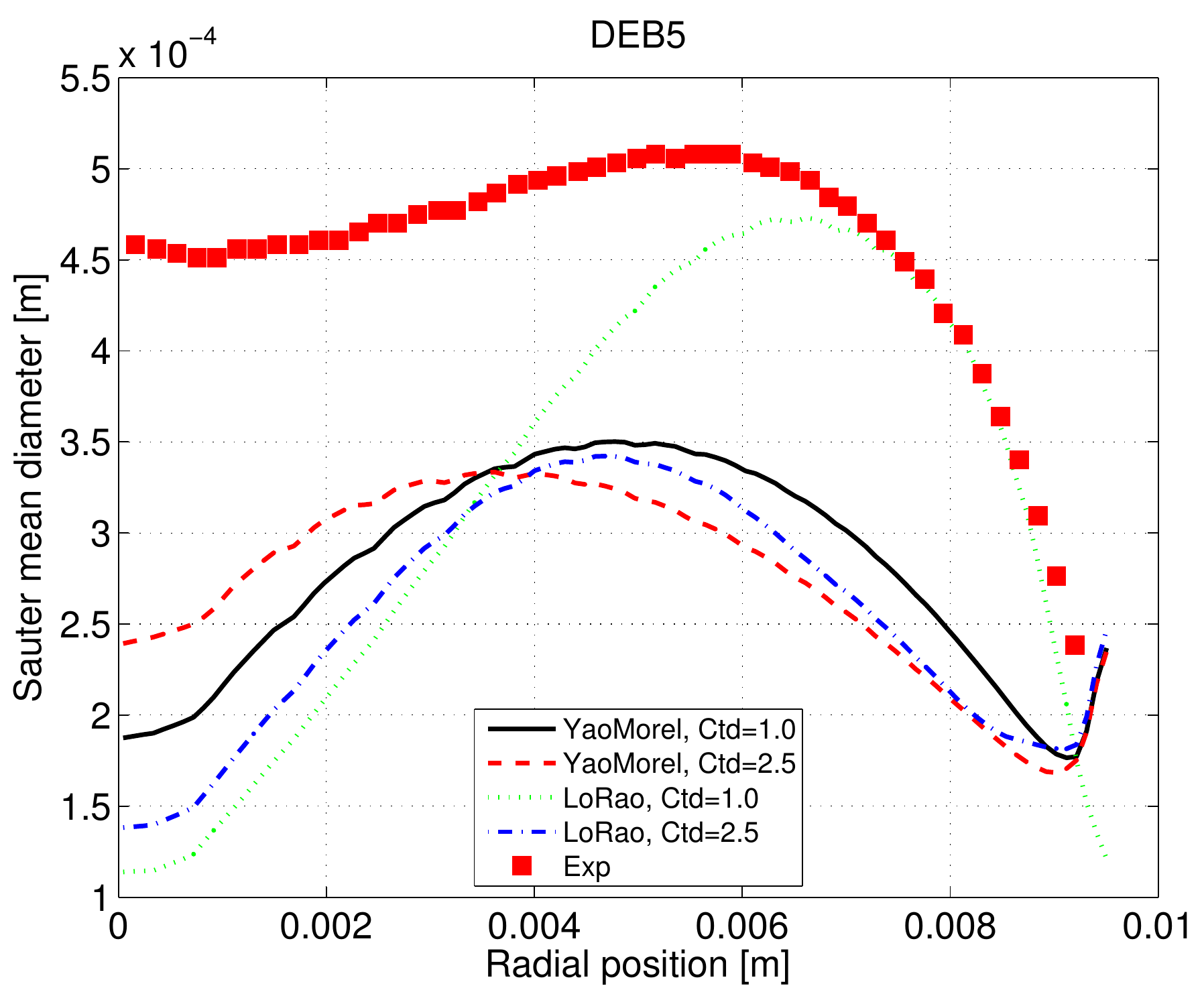}
    \vspace*{0cm}\hspace*{0cm}\label{fig3:a}
}
\subfloat[]
{
    \vspace*{0cm}\hspace*{0cm}
    \epsfxsize=195pt\epsfbox{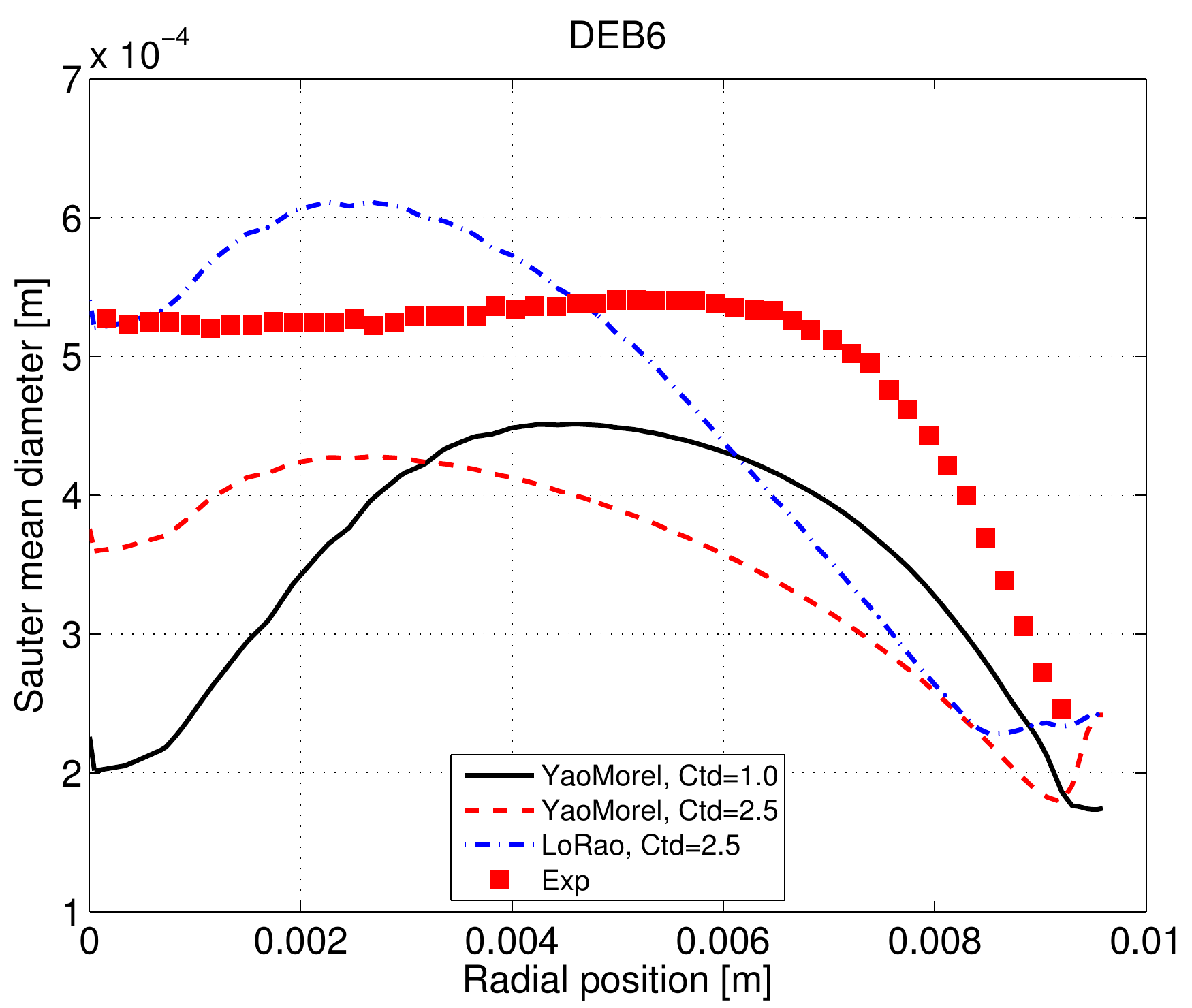}
    \vspace*{0cm}\hspace*{0cm}\label{fig3:b}
}
\caption{Comparison between the DEBORA experiment and calculation results: Radial Sauter mean diameter}
\label{fig3}
\end{figure}

\begin{figure}
\subfloat[]
{
    \vspace*{0cm}\hspace*{-1cm}
    \epsfxsize=200pt\epsfbox{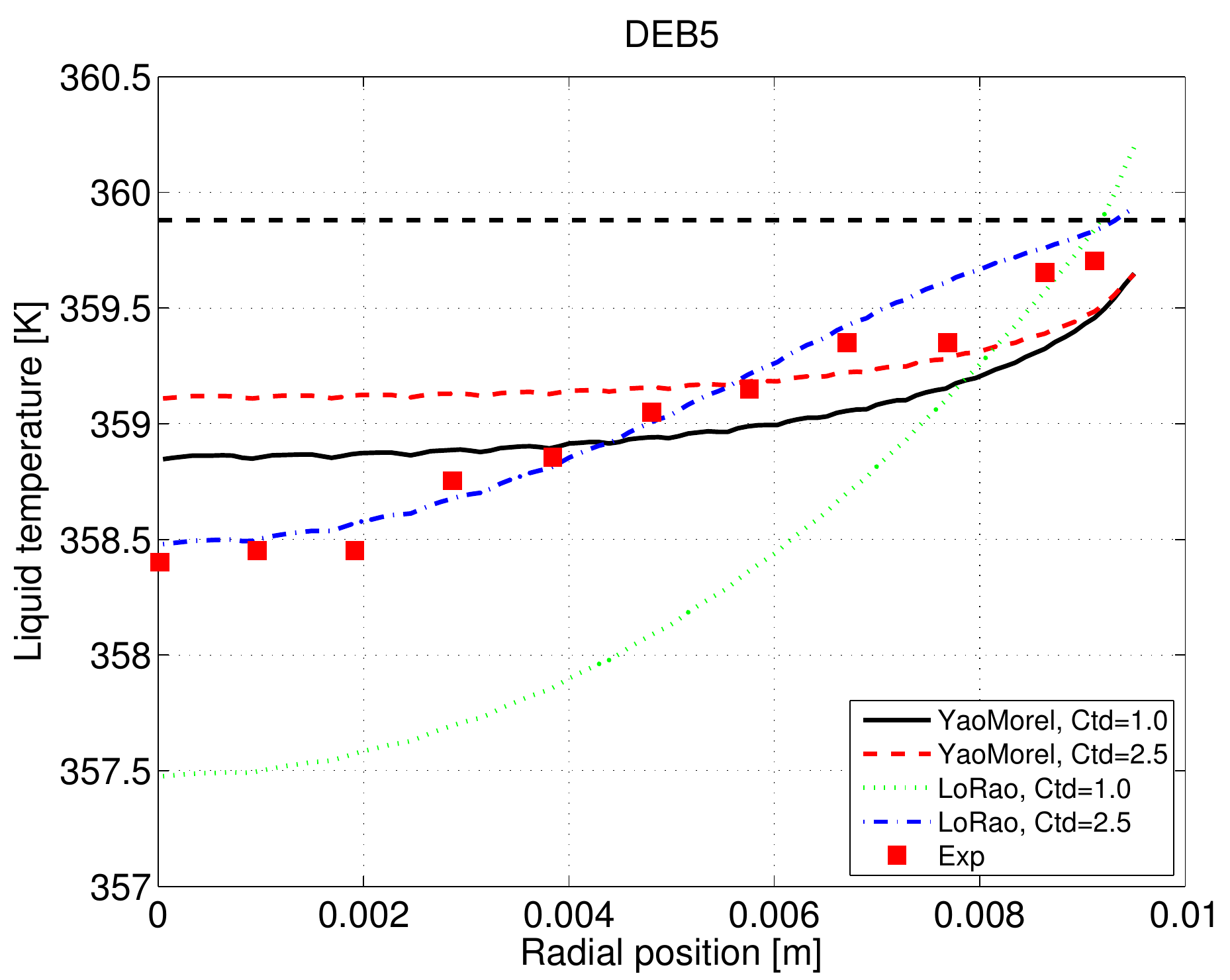}
    \vspace*{0cm}\hspace*{0cm}\label{fig4:a}
}
\subfloat[]
{
    \vspace*{0cm}\hspace*{0cm}
    \epsfxsize=200pt\epsfbox{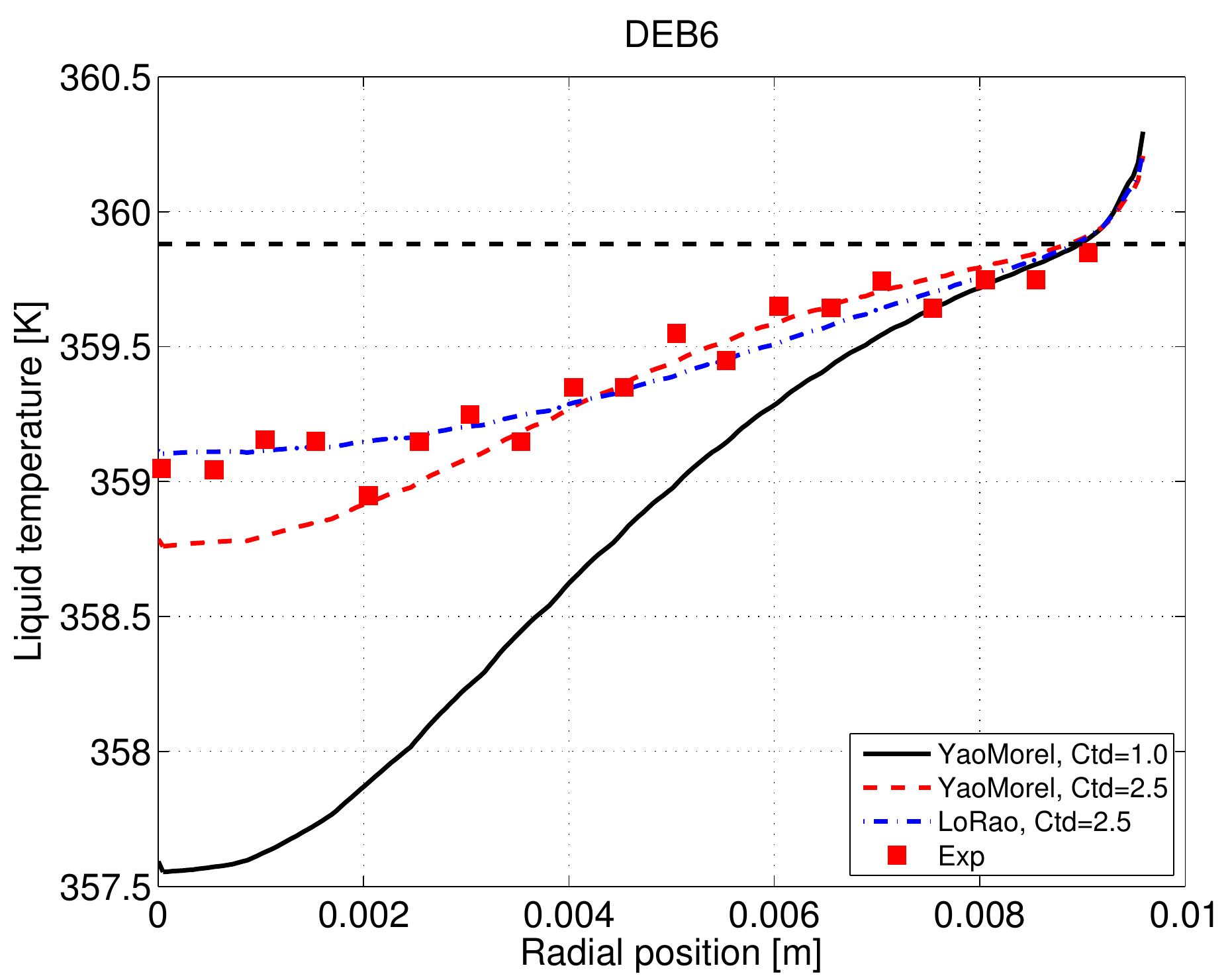}
    \vspace*{0cm}\hspace*{0cm}\label{fig4:b}
}
\caption{Comparison between the DEBORA experiment and calculation results: Radial liquid temperature}
\label{fig4}
\end{figure}

Figures \ref{fig2} - \ref{fig4} show the comparison between the measured and predicted results of DEBORA experiment. Two sets of breakup and coalescence models were tested in our simulation. One should notice that \cite{Yao} breakup and coalescence model is used together with their turbulence modeling and \cite{Lo} breakup and coalescence model together with the standard $k-\epsilon$ model. In addition, the sensitivity of turbulent dispersion force coefficient was tested here. The suggested value of $C_{td}$ is usually in the region [0.1, 1.0] for bubbly flow. However $C_{td}=1.0$ is not sufficient enough to push the evaporation bubbles away from the surface, leading to an accumulation of void fraction near the wall, as shown in Fig. \ref{fig2:a}. Due to that the local void fraction close to the wall may reach too high levels (above 0.74) exceeding the limits of the applicability of the present bubbly flow model. That is why we could not do the simulation with \cite{Lo} model together with $C_{td}=1.0$ in case of DEB6, as shown in Fig. \ref{fig2:b}.

%The major problem in the simulation is the prediction of bubble diameter. As we can see from the experiment data, the bubble grows fast after its departure from the wall and soon reaches its stable diameter. The growth of bubble size could be explained by the contribution of coalescence. However the stable bubble size cannot be explained by the existing models. In our simulation, the condensation of bubble in the bulk could result in the shrinkage of its size, as show in Fig. \ref{fig3}. The prediction of liquid temperature as shown in Fig. \ref{fig4} is fairly well, which is consistent with the energy conservation.

In general, a quite satisfactory agreement between the measured and the calculated void fraction distribution has been obtained. In particular, Fig. \ref{fig2:b} reveals that significant improvement in over-all accuracy can be obtained by choosing the turbulence dispersion force coefficient in the range between 1.0 and 2.5. The accuracy of prediction of bubble size is, however, not satisfactory. As shown in Fig. \ref{fig3:a}, the bubble size is significantly underestimated in the observation part of the test section. This could be caused by underestimation of the bubble coalescence rate in this region. The results indicate that more work is needed to improve the interfacial area transport models. Figure \ref{fig4} shows a very good agreement between predicted and measured radial temperature distributions for both cases.

\section{Conclusion}
A two-fluid boiling flow model has been implemented into the OpenFOAM solver and validated against the Bartolomej and the DEBORA experimental data. The model includes the closure relationships for the heat transfer and phase change for bubbles moving in a subcooled liquid. Bubble size is predicted from the interfacial area concentration transport equations, including the source and sink terms resulting from the bubble coalescence and breakup, nucleation at walls as well as phase change induced source term. The present model has been validated against measurements performed in a vertical upward flow in a heated pipe. The prediction of void fraction as well as the liquid temperature profile could be done with quite satisfactory accuracy. The accuracy of prediction of the bubble size distribution is found quite low, indicating that still more work is needed to improve the interfacial area transport models.

\section{Acknowledgments}
Financial supports from NORTHNET, as well as support from the Swedish National Infrastructure for Computing are gratefully acknowledged.
\cleardoublepage

\begin{supertabular}{ll}

\multicolumn{2}{l}{\textbf{Nomenclature}}\\
&\\

$A$& area fraction\\
$a_i$    &    interfacial area concentration, m$^{-1}$\\
$C$    &    interfacial force coefficient\\
$c_{li}$& heat transfer coefficient given by Eqn. \ref{cli}, kg$\cdot$m$^{-2}\cdot$s$^{-1}$\\
$c_{p}$ &       specific heat, J$\cdot$kg$^{-1}\cdot$K$^{-1}$\\
$c_{vi}$& heat transfer coefficient given by Eqn. \ref{cvi}, kg$\cdot$m$^{-3}\cdot$s$^{-1}$\\
$D_S$   &Sauter mean diameter, m\\
$d$&     diameter, m \\
$d_h$   &hydraulic diameter, m\\
Eo      &Eotvos number \\
%$F$&force\\
$f$     &   bubble departure frequency, s$^{-1}$ \\
$G$&turbulent production,  kg$\cdot$m$^{-1}\cdot$s$^{-3}$ or mass flow rate, kg$\cdot$m$^{-2}\cdot$s$^{-1}$ \\
$g$& gravity constant, m$\cdot$s$^{-2}$\\
$h$& enthalpy, J$\cdot$kg$^{-1}$ \\
$h_{fc}$& single phase convective heat transfer coefficient, W$\cdot$m$^{-2}\cdot$K$^{-1}$\\
$h_{fg}$&   latent heat, J$\cdot$kg$^{-1}$  \\
$h_{q}$&  quenching heat transfer coefficient, W$\cdot$m$^{-2}\cdot$K$^{-1}$\\
Ja&  Jacob number\\
$k$  & turbulent kinetic energy, m$^2\cdot$s$^{-2}$ \\
$\mathbf M$& interfacial momentum transfer rate, kg$\cdot$m$^{-2}\cdot$s$^{-2}$ \\
$N^{\prime\prime}$&   active nucleation site density, m$^{-2}$ \\
Nu  &  Nusselt number\\
$\mathbf n_w$& unit vector normal to wall \\
Pe&P\'{e}clet number\\
Pr  & Prandtl number \\
$p$ & pressure, Pa\\
$ q^{\prime\prime},{\mathbf q}^{\prime\prime}$&heat flux, W$\cdot$m$^{-2}$ \\
$ q^{\prime\prime\prime}$&heat flow rate per unit volume, W$\cdot$m$^{-3}$ \\
Re  &    Reynolds number\\
%$r$ & radius\\
$T$&temperature, K\\
$t$       &  time, s\\
$\mathbf U$&velocity, m$\cdot$s$^{-1}$\\
$u_\tau$& friction velocity, m$\cdot$s$^{-1}$\\
We& Weber number\\
&\\
\multicolumn{2}{l}{\emph{Greek letters}}\\
$\alpha$ & void fraction\\
$\beta$ & void fraction for continuous phase\\
$\epsilon$&turbulent dissipation rate, m$^2\cdot$s$^{-3}$\\
$\Gamma$&rate of phase change, kg$\cdot$m$^{-3}\cdot$s$^{-1}$\\
$\kappa$    &           thermal diffusivity, m$^2\cdot$s$^{-1}$\\
$\lambda$ & thermal conductivity, W$\cdot$m$^{-1}\cdot$K$^{-1}$\\
$\mu$      & dynamic viscosity, kg$\cdot$m$^{-1}\cdot$s$^{-1}$\\
$\nu$   &   kinematic viscosity, m$^{2}\cdot$s$^{-1}$ \\
$\psi$  &  factor depending on bubble shape \\
$\rho$ &    density, kg$\cdot$m$^{-3}$\\
$\sigma$ &  interfacial tension, N$\cdot$m$^{-1}$ \\
$\bm \tau$     &  stress tensor, N$\cdot$m$^{-2}$\\
$\tau_w$&wall shear stress, N$\cdot$m$^{-2}$\\
%$\theta$    &   contact angle\\
&\\

\multicolumn{2}{l}{\emph{Superscripts}}\\
eff & effective\\
$d$ & drag\\
$l$ & lift\\
$t$ & turbulence\\
$td$ & turbulent dispersion\\
$vm$ & virtual mass \\
$w$ & wall \\
$wl$ & wall lubrication\\

&\\

\multicolumn{2}{l}{\emph{Subscripts}}\\
$1\Phi$&single phase\\
BB& bubble breakup\\
BC& bubble coalescence\\
$b$& bubble\\
$c$& convection\\
$fc$  & single phase forced convection \\
%$d$  & bubble departure \\
%$ev$& evaporation\\
$i$& interphase\\
$k$ & phase\\
$l$& liquid\\
$lo$& lift-off\\
%$n$& normal or $n$th\\
NUC& nucleation \\
%$p$&pressure\\
$q$&quenching\\
$r$&relative\\
ref&reference\\
sat&saturation\\
$v$&vapor\\
$w$& wall\\

\end{supertabular}

\bibliographystyle{elsarticle-harv}
\bibliography{refs}

\end{document}